\documentclass[reprint,superscriptaddress, amsmath,amssymb,aps]{revtex4-1}
\usepackage{graphicx}
\usepackage{dcolumn}
\usepackage{bm}

\usepackage[dvipdfmx]{color}

\begin{document}

\title{Universal Equation of State Describes Osmotic Pressure throughout Gelation Process}

\author{Takashi~Yasuda}
\thanks{These authors contributed equally: T.~Yasuda, N.~Sakumichi}
\affiliation{Department of Bioengineering, The University of Tokyo, 7-3-1 Hongo, Bunkyo-ku, Tokyo, Japan.}
\author{Naoyuki~Sakumichi}
\email[Correspondence should be addressed to N.~Sakumichi or T.~Sakai: ]{sakumichi@tetrapod.t.u-tokyo.ac.jp}
\affiliation{Department of Bioengineering, The University of Tokyo, 7-3-1 Hongo, Bunkyo-ku, Tokyo, Japan.}
\author{Ung-il~Chung}
\affiliation{Department of Bioengineering, The University of Tokyo, 7-3-1 Hongo, Bunkyo-ku, Tokyo, Japan.}
\author{Takamasa~Sakai}
\email{sakai@tetrapod.t.u-tokyo.ac.jp}
\affiliation{Department of Bioengineering, The University of Tokyo, 7-3-1 Hongo, Bunkyo-ku, Tokyo, Japan.}
\date{\today}

\begin{abstract}
The equation of state of the osmotic pressure for linear-polymer solutions in good solvents is universally described by a scaling function.
We experimentally measure the osmotic pressure of the gelation process via osmotic deswelling.
We find that the same scaling function for linear-polymer solutions also describes the osmotic pressure throughout the gelation process involving both the sol and gel states.
Furthermore, we reveal that the osmotic pressure of polymer gels is universally governed by the semidilute scaling law of linear-polymer solutions.
\end{abstract}

\maketitle

Flexible linear polymers in good solvents provide not only the basis of polymer physics \cite{flory1953principles,de1979scaling}, but also a remarkable example of the notion of universality of critical phenomena in statistical physics \cite{de1979scaling,oono1985statistical}.
Their macroscopic collective properties are independent of the microscopic details of the system and are described by a small number of parameters, because of the great length of polymer chains.
Such systems belong to the $O(n)$-symmetric universality classes ($n=1,2,3$ corresponding to the Ising, \textit{XY}, and Heisenberg classes, respectively) found in many systems, ranging from those of soft and hard condensed-matter physics to high-energy physics \cite{pelissetto2002critical}.
The above linear-polymer solutions correspond to the limit of $n\to 0$ (self-avoiding walks) in three dimensions \cite{de1979scaling,pelissetto2002critical}, for which the critical exponent (the excluded volume parameter) $\nu \simeq 0.588$ can be computed using three independent methods: Monte Carlo simulations \cite{clisby2010accurate,clisby2016high}, the $\epsilon$-expansion method \cite{kompaniets2017minimally}, and the conformal bootstrap method \cite{shimada2016fractal,hikami2018conformal}.
Furthermore, not only the critical exponents but also the asymptotic scaling functions themselves can be experimentally measured, such as the osmotic pressure \cite{noda1981thermodynamic,higo1983osmotic} and the correlation lengths of density fluctuations \cite{wiltzius1983universality}.\\

We focus on the equation of state (EOS) of osmotic pressure for (electrically neutral) flexible linear polymers in good solvents, which is universally described by the following scaling function
\cite{noda1981thermodynamic,higo1983osmotic,des1975lagrangian,des1982osmotic,ohta1982conformation,ohta1983theory}:
\begin{equation}
\hat{\Pi} = f\left({\hat{c}}\right),
\label{eq:EOS}
\end{equation}
where $\hat{\Pi} \equiv \Pi M/(cRT)$ is the reduced osmotic pressure, and $\hat{c}\equiv c/c^*$ is the reduced polymer concentration normalized by the overlap concentration $c^{*}\equiv1/(A_2 M)$.
Here, $M$, $R$, $T$, and $A_2$ are the molar mass, gas constant, absolute temperature, and the second virial coefficient, respectively.
The above definition of $c^{*}$ is proportional \cite{burchard1999solution} to the conventional definition of the overlap concentration $c^*_g \equiv 3M/(4\pi N_A R_g^3)$, at which the polymer chains begin to overlap to fill the space.
Here, $N_A$ and $R_g$ are the Avogadro constant and the gyration radius of the polymer chain, respectively.\\

\begin{figure}[b!]
\centering
\includegraphics[width=\linewidth]{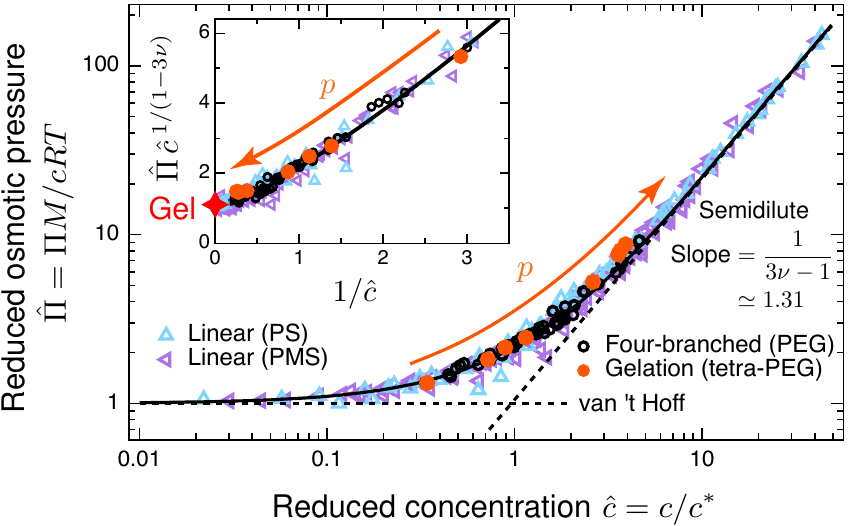}
\caption{
Universal EOS of osmotic pressure for polymer solutions and gels in a good solvent.
Main image shows the $\hat{c}$ dependence of $\hat{\Pi}$ in a log-log plot. Inset: the $\hat{c}^{-1}$ dependence of $\hat{\Pi}\,\hat{c}^{1/(1-3\nu)}$.
The triangles represent two kinds of linear polymers [poly(styrene) (PS), where $M=51$--$1900$ kg$/$mol \cite{higo1983osmotic}, and poly($\alpha$-methylstyrene) (PMS), where $M= 70.8$--$1820$ kg$/$mol \cite{noda1981thermodynamic}] in toluene solutions.
The triangles converge to the universal EOS~(\ref{eq:EOS}) (black solid curve), which is asymptotic to the van 't Hoff law ($\hat{\Pi}=1$) as $\hat{c}\to 0$ and to the scaling law in Eq.~(\ref{eq:scaling}) as $\hat{c}\to \infty$ (black dotted lines).
The black circles represent the four-branched polymer [poly(ethylene glycol) (PEG)] solutions of $M=10$ and $40$~kg$/$mol.
The orange-filled circles represent the sol samples that emulate the gelation process with varying degrees of connectivity ($p=0, 0.1, \dots, 0.5$) at a constant concentration ($c=20$ g$/$L).
The red star in the inset corresponds to the gel samples.
}
\label{fig:EOS}
\end{figure}

For branched polymer solutions, it was reported that each EOS of regular star polymers with up to 18 arms exhibited only minor differences from the universal EOS~(\ref{eq:EOS}) of linear polymers
\cite{higo1983osmotic,adam1991concentration,merkle1993osmotic,burchard1999solution}.
Here, $\hat{c}\equiv c/c^*$ is the only universal scaling parameter (up to multiplication by a constant) \cite{burchard1999solution}.
Hence, $c/c^*_{g}$ is not a universal scaling parameter because $c_{g}^{*}/c^*=3\sqrt{\pi}\Psi^{*}$ includes the interpenetration factor $\Psi^{*}$, which is nonuniversal for a number of arms (e.g., $\Psi^{*}\simeq 0.24$ and $0.44$ for linear and four-branched polymer solutions, respectively \cite{rubio1996monte,okumoto1998excluded}).
Figure~\ref{fig:EOS} demonstrates that the different kinds of linear-polymer solutions and four-branched polymer solutions converge to a universal EOS~(\ref{eq:EOS}).
In the dilute regime ($0\leq\hat{c}<1$), each molecular chain is isolated to sufficiently describe the universal EOS~(\ref{eq:EOS}) through virial expansion \cite{flory1953principles},
\begin{equation}
\hat{\Pi} = f\left({\hat{c}}\right) = 1 + \hat{c} + \gamma\,\hat{c}^{2}+ \cdots 
\quad(\mathrm{for}\,\,\, 0\leq\hat{c}<1),
\label{eq:virial}
\end{equation}
where $\gamma \simeq 0.25$ is the dimensionless virial ratio \cite{flory1953principles,noda1981thermodynamic}.
In the semidilute regime ($\hat{c}>1$), the molecular chains interpenetrate, and the universal EOS~(\ref{eq:EOS}) is asymptotic to the semidilute scaling law \cite{des1975lagrangian,de1979scaling}
\begin{equation}
\hat{\Pi} = f\left({\hat{c}}\right) \simeq K\hat{c}^{\frac{1}{3\nu -1}}
\qquad(\mathrm{for}\,\,\, \hat{c}\gg 1),
\label{eq:scaling}
\end{equation}
where $K\simeq 1.1$ and $1/(3\nu -1)\simeq 1.31$ because $\nu\simeq0.588$.\\

In this Letter, we experimentally investigate the osmotic pressure of polymer gels throughout the gelation process, which involves both the sol and gel states.
We measured the osmotic pressure via osmotic deswelling in external polymer solutions \cite{bastide1981osmotic,horkay1986studies,horkay2000osmotic}.
Our findings are summarized in Fig.~\ref{fig:EOS}; 
the universal EOS~(\ref{eq:EOS}) describes the osmotic pressure of both the sol (orange-filled circles) and gel (red star) states with only minor variations, although each system during the gelation process comprises highly branched polymer networks.
When gelation proceeds at a constant concentration $c$,
the average molar mass $M$ increases and $c^{*}$ decreases.
Thus, both $\hat{\Pi}$ and $\hat{c}$ continuously increase along the universal EOS~(\ref{eq:EOS}) in the sol state.
After the gelation (i.e., the sol-gel transition), because polymer gels correspond to $M\to\infty$ and $c^{*}\to 0$, both $\hat{\Pi}$ and $\hat{c}$ diverge to infinity in the gel state.
According to the semidilute scaling law described by Eq.~(\ref{eq:scaling}), $\hat{\Pi}\,\hat{c}^{1/(1-3\nu)}$ remains constant in the gel state (red star in the inset of Fig.~\ref{fig:EOS}).
Our findings, which elucidate the universal laws governing osmotic pressure, are not only conceptually important for statistical physics, but also practically useful for soft-matter physics.
These results are essential for the applications of polymer solutions and polymer gels, which can swell by imbibing solvents.\\

\begin{figure}[t!]
\centering
\includegraphics[width=\linewidth]{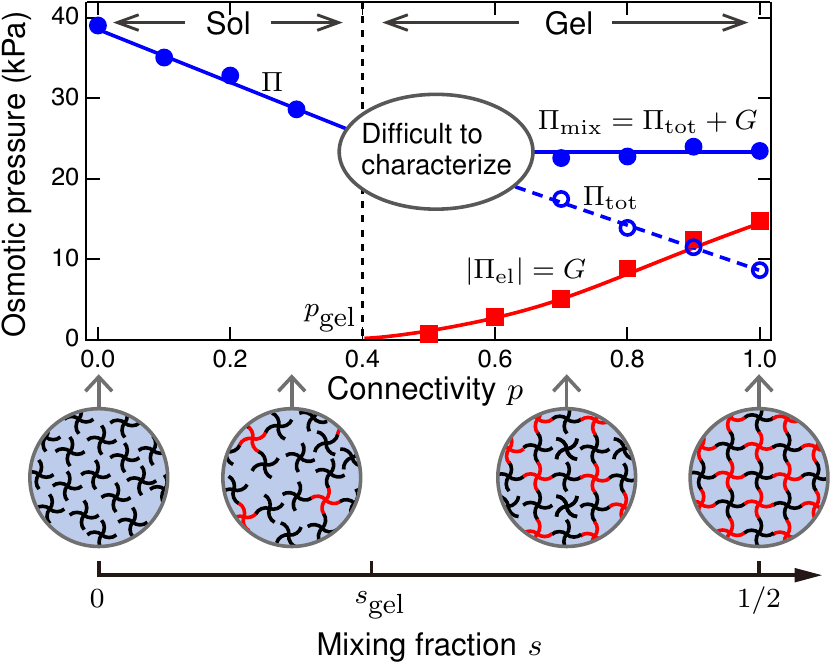}
\caption{
Osmotic pressure of the samples emulating gelation process.
The samples were prepared at a constant polymer concentration ($c=60$ g$/$L) and molar mass of precursors ($M=10$ kg$/$mol).
By measuring $\Pi_\mathrm{tot}$ and $G$, we obtained $\Pi_\mathrm{mix}=\Pi_\mathrm{tot}+G$ in the gel state.
As the connectivity $p$ increases, $\Pi$ and $\Pi_\mathrm{tot}$ decrease, and $\Pi_{\mathrm{mix}}$ remains constant (blue curves).
After gelation ($p_{\mathrm{gel}}\leq p\leq1$), the elasticity (red curve) increases.
Here, $p$ ($0\leq p\leq1$) is controlled by nonstoichiometrically mixing two types of precursors in an \textit{AB}-type polymerization system.
Gel samples with a low connectivity ($p_{\mathrm{gel}} \leq p < 0.7$) were difficult to characterize, because of the outflow of small polymer clusters.
}
\label{fig:gelation}
\end{figure}

To statically emulate the gelation process, we nonstoichiometrically tuned the mixing fractions $s$ ($0\leq s\leq 1/2$) of two types of precursor solutions in an \textit{AB}-type polymerization system (schematics in Fig.~\ref{fig:gelation}).
Here, $s$ is the molar fraction of the minor precursors to all precursors.
We define the connectivity $p$ ($0\leq p\leq1$) as the fraction of reacted terminal functional groups, assuming the reaction is completed.
By tuning $s$ in accordance with $p = 2s$ \cite{sakai2016sol,yoshikawa2019connectivity}, we can obtain a desired $p$.
Before gelation ($0\leq p <p_\mathrm{gel}$), polymer chains crosslink to form a polydisperse mixture of highly branched polymers with increases in the average molar mass $M$.
After gelation ($p_\mathrm{gel}\leq p\leq 1$), these polymer networks cross-link to complete the reaction as the elasticity increases.\\

Based on our findings displayed in Fig.~\ref{fig:EOS}, we illustrate the ``nonreduced'' osmotic pressure $\Pi$ during the gelation process  in Fig.~\ref{fig:gelation}.
Unlike the sol state, the gel state has elastic contributions to the total swelling pressure ($\Pi_{\mathrm{tot}}$).
According to Flory and Rehner \cite{flory1943jr}, $\Pi_{\mathrm{tot}}$ consists of two separate contributions as
$\Pi_{\mathrm{tot}}=\Pi_{\mathrm{mix}}+\Pi_{\mathrm{el}}$,
where $\Pi_{\mathrm{mix}}$ and $\Pi_{\mathrm{el}}$ are the mixing and elastic contributions, respectively.
We regard $\Pi_{\mathrm{mix}}$ as the osmotic pressure in the gel state, because $\Pi_{\mathrm{mix}}$ corresponds to the osmotic pressure in the sol state $\Pi$.
As the connectivity $p$ increases at a constant concentration $c$, $\Pi$ in the sol state decreases because the chemical reaction decreases the number density of the molecules.
After gelation, $\Pi_{\mathrm{mix}}$ reaches a constant; polymer gels are always in a semidilute regime with an infinite molar mass.\\

\textit{Materials and methods}.---For the model system of \textit{AB}-type polymerization in gelation, we used a tetra-arm poly(ethylene glycol) (tetra-PEG) gel that is synthesized by the \textit{AB}-type cross-end coupling of two tetra-PEG units of equal size \cite{sakai2008design}. 
Each end of the tetra-PEG is modified with mutually reactive maleimide (tetra-PEG MA) and thiol (tetra-PEG SH).
We dissolved tetra-PEG MA and tetra-PEG SH (Nippon Oil \& Fat Corporation) in a phosphate-citrate buffer with an ionic strength and \textit{p}H of $200$ mM and $3.8$, respectively.
For gelation, we mixed these solutions with equal molar masses $M$ and equal concentrations $c$ in various mixing fractions $s$.
We kept each sample in an enclosed space to maintain humid conditions at room temperature ($T\simeq 298$ K) to allow completion of the reaction.\\

\begin{figure}[t!]
\centering
\includegraphics[width=\linewidth]{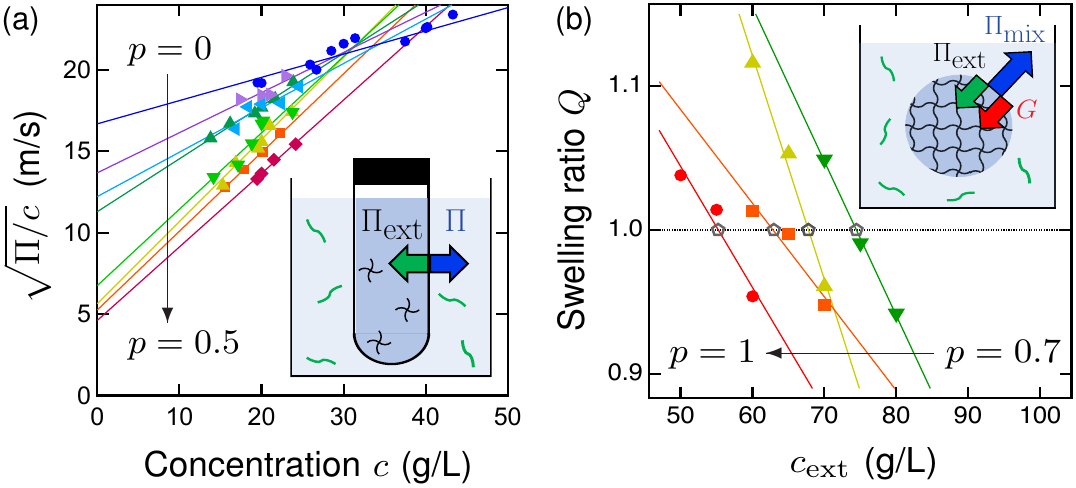}
\caption{
Osmotic deswelling in external polymer solutions, used to measure $\Pi$ and $\Pi_{\mathrm{mix}}$, in (a) sol and (b) gel samples, respectively.
For each plot, the precursors are $M=10$~kg$/$mol.
Each line is the least-squares fit to the data for each $p$.
(a) Square-root plots of $\Pi$ of sol samples on $c_{0}=20$ g$/$L for $p=0, 0.1, 0.2, 0.25, 0.3, 0.35, 0.4,$ and $0.5$.
We immersed samples with a microdialyzer in external polymer (PVP) solutions.
We can determine $\Pi$, because $\Pi = \Pi_{\mathrm{ext}}$ at equilibrium.
(b) Equilibrium swelling ratio $Q$ of gel samples on $c_{0}=60$ g$/$L for $p=0.7,0.8,0.9,$ and $1$ in the external polymer (PVP) solutions.
We directly immersed samples in external solutions of various concentrations $c_\mathrm{ext}$.
We can determine $\Pi_{\mathrm{mix}}$, because $\Pi_{\mathrm{mix}}=\Pi_{\mathrm{ext}}+G$ at equilibrium.
}
\label{fig:measurement}
\end{figure}

We prepared the four-branched polymer (precursor) solutions ($p=0$) by dissolving tetra-PEG MA with molar masses of $M=10$ and $40$~kg$/$mol and initial concentrations $c_{0} = 20$--$120$ g$/$L.
Herein, we define the polymer concentration ($c_{0}$ and $c$) as the precursor weight divided by the solvent volume, rather than by the solution volume, to extend the universality of the EOS~(\ref{eq:EOS}) to higher concentrations (see Supplemental Material, Sec.~S1).
We prepared the sol and gel samples that emulate the gelation process by dissolving precursors with $M=10$~kg$/$mol.
For $c_{0} = 20$ g$/$L, we set $p=2s=0.1, 0.2, 0.25, 0.3, 0.35, 0.4,$ and $0.5$ (sol samples).
For $c_{0}=40, 60, 80,$ and $120$ g$/$L, we set $p=2s=0.1, 0.2,$ and  $0.3$ (sol samples) and $0.7, 0.8, 0.9,$ and $1$ (gel samples).
Section~S2 of the Supplemental Material describes the determination of these measurement ranges.\\

We measured the osmotic pressures in the sol samples $\Pi$, using controlled aqueous poly(vinylpyrrolidone) (PVP, K90, Sigma Aldrich) solutions whose concentration dependence of osmotic pressure $\Pi_{\mathrm{ext}}$ was measured by Vink \cite{vink1971precision} (Supplemental Material, Sec.~S3).
As shown in the schematic in Fig.~\ref{fig:measurement}(a), each solution sample was placed in a microdialyzer (MD300, Scienova) that had a semipermeable membrane with a mesh size of $3.5$ kDa.
We immersed each dialyzer in an aqueous polymer (PVP) solution at a certain concentration $c_\mathrm{ext}$ with stirring.
Subsequently, each system achieved equilibrium at $\Pi = \Pi_{\mathrm{ext}}$.
(The achievement of swelling equilibrium was assured.
See Supplemental Material, Sec.~S4.)
At that time, each solution sample was changed in weight and concentration from its initial to equilibrium states, as represented by $W_0 \to W$ and $c_0 \to c$, respectively.
Assuming a constant weight density and small deformation for the sample, we calculate $c$ as $c=c_{0}/Q$, where $Q=W/W_{0}$ is the equilibrium swelling ratio.
In examining the gelation process (e.g., Fig.~\ref{fig:gelation}), we evaluated $\Pi$ of the ``as-prepared'' (i.e., $Q=1$) sol samples at equal concentrations $c=c_0$ with various values of $p$.
Measuring $Q$ for various $c_\mathrm{ext}$ and interpolating the $c_\mathrm{ext}$ dependence of $Q$ as a linear function, we determined $c_\mathrm{ext}$ and $\Pi_\mathrm{ext}$ such that each sol sample maintained its weight ($Q=1$) and concentration ($c=c_{0}$).\\

To evaluate the parameters $M$ and $c^{*}$ from $\Pi = \Pi (c)$, which were measured at each $p$, we used square-root plots \cite{flory1953principles}, as shown in Fig.~\ref{fig:measurement}(a).
From the virial expansion~(\ref{eq:virial}), we have
$\hat{\Pi}
 = \left[1+\hat{c}/2
 + \left(\gamma-1/4 \right)\hat{c}^2/2
 \right]^2
 + O\left(\hat{c}^3\right)
$.
Together with $\gamma \simeq 1/4$ (Supplemental Material, Sec.~S5) for certain few-branched polymer solutions, we have
$\sqrt{\Pi/c}
\simeq \sqrt{RT/M}
\left[1+c/(2c^{*})\right]$
for small $c/c^{*}$.
Thus, the intercept and slope of each fitting line in Fig.~\ref{fig:measurement}(a) give $M$ and $c^{*}$, respectively, for each $p$.
The obtained $M$ and $c^{*}$ values are consistent with the scaling prediction of $c^{*} \sim M^{1-3\nu}$ (Supplemental Material, Sec.~S6).\\

We measured the osmotic pressure in the as-prepared gel states $\Pi_\mathrm{mix}$ via osmotic deswelling.
As shown in the schematic in Fig.~\ref{fig:measurement}(b), we immersed each gel sample directly in the external aqueous polymer (PVP) solutions of various concentrations $c_\mathrm{ext}$, because the surfaces of the gels function as semipermeable membranes.
Subsequently, each gel sample swelled or deswelled toward equilibrium at $\Pi_{\mathrm{mix}}+\Pi_{\mathrm{el}}=\Pi_{\mathrm{ext}}$, changing its weight and concentration from the as-prepared to equilibrium states as represented by $W_0 \to W$ and $c_0 \to c$, respectively.
The equilibrium swelling ratio $Q=W/W_{0}$ was measured and interpolated as a linear function of $c_\mathrm{ext}$ for each gel sample [examples are given in Fig.~\ref{fig:measurement}(b)].
Using the $c_\mathrm{ext}$ dependence of $Q$, we evaluated $c_\mathrm{ext}$ and $\Pi_\mathrm{ext}$ such that each gel sample maintained its weight ($Q = 1$) and concentration ($c = c_{0}$).
(This method is the same as the above method to determine $\Pi$ of the as-prepared sol samples.)
Assuming $\Pi_{\mathrm{el}} = -G$ \cite{james1949simple}, we evaluated $\Pi_{\mathrm{mix}} = \Pi_{\mathrm{ext}} + G$ for each as-prepared gel sample, where $G$ is the shear modulus as measured by rheometry (Supplemental Material, Sec.~S7).\\

\begin{figure}[t!]
\centering
\includegraphics[width=\linewidth]{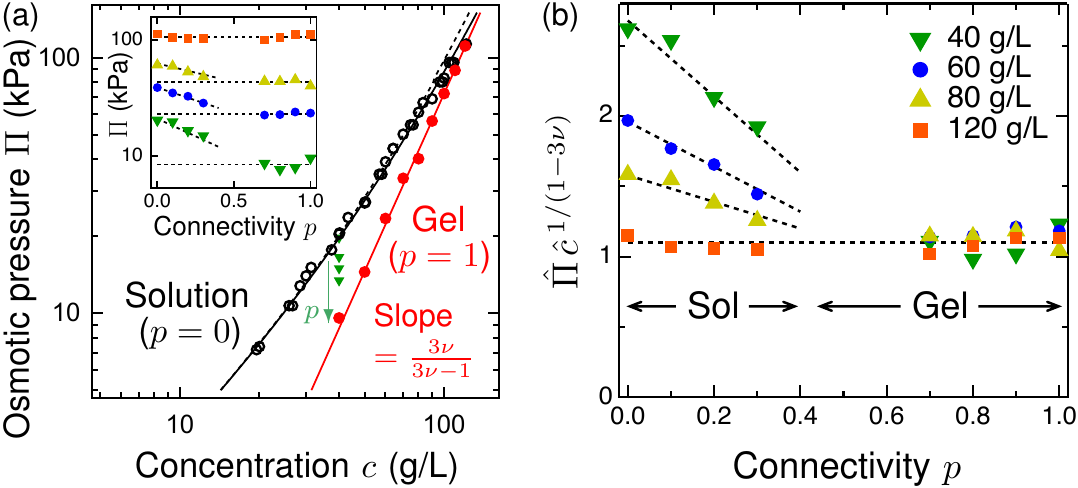}
\caption{
Osmotic pressure during gelation process.
The molar mass of precursors is $M=10$ kg$/$mol, corresponding to the overlap concentration $c^*\simeq58$~g$/$L at $p=0$.
(a) Osmotic pressure in the unreacted four-branched polymer solutions (black circles) and in the reaction-completed polymer gels (red-filled circles).
The former and latter agree with the universal EOS~(\ref{eq:EOS}) (black curve) and with the semidilute scaling law $\Pi\propto c^{3\nu/(3\nu -1)}$ (red line), respectively.
Here, $3\nu/(3\nu -1)\simeq 2.31$ because $\nu \simeq 0.588$.
The black dotted curve is the virial expansion~(\ref{eq:virial}) up to the third-order terms.
As $p$ increases (green triangles), $\Pi$ decreases in the sol state ($0\leq p<p_\mathrm{gel}$) and remains constant in the gel state ($p_\mathrm{gel}<p\leq1$).
Inset: osmotic pressure during the gelation process at a constant polymer concentration $c=c_{0}=40, 60, 80,$ and $120$~g$/$L.
The green triangles ($c=40$~g$/$L) are the same as those in the main panel.
The blue circles ($c=60$~g$/$L) are used in Fig.~\ref{fig:gelation}.
(b) Connectivity ($p$) dependence of $\hat{\Pi}\,\hat{c}^{1/(1-3\nu)}$.
The symbols and data are the same as those in the inset of (a).
In the gel state, $\hat{\Pi}\,\hat{c}^{1/(1-3\nu)}$ converge to the universal value $K\simeq 1.1$, which is independent of $p$ and $c$.
}
\label{fig:result}
\end{figure}

\textit{Results and analysis}.---The main panel in Fig.~\ref{fig:result}(a) shows the $c$ dependence of the osmotic pressure in both the unreacted four-branched polymer solutions ($p = 0$) and the reaction-completed polymer gels ($p = 1$).
The experimental results for the former and latter agree with the universal EOS~(\ref{eq:EOS}) for \textit{linear}-polymer solutions and with the semidilute scaling law $\Pi\propto c^{3\nu/(3\nu -1)}$, respectively, in the wide concentration range $c$.
With an increase in $c$, $\Pi$ in the polymer solutions (black curve) is asymptotic to $\Pi_{\mathrm{mix}}$ in the polymer gels (red line).
This asymptotic relationship suggests that $\Pi_{\mathrm{mix}}$ of polymer gels is governed by the semidilute scaling law described in Eq.~(\ref{eq:scaling}) with $K\simeq 1.1$ for polymer solutions.\\

The inset in Fig.~\ref{fig:result}(a) shows the $p$ dependence of $\Pi$ and $\Pi_\mathrm{mix}$ throughout the gelation process ($0\leq p \leq1$). 
In the sol state ($0\leq p<p_\mathrm{gel}$), $\Pi$ decreases as $p$ increases, because the average molar mass $M$ increases.
As $c$ increases, the extent of the decrease in the osmotic pressure itself decreases.
In particular, for $c=120$ g$/$L, $\Pi$ and $\Pi_\mathrm{mix}$ are constant throughout the gelation process ($0\leq p \leq1$), because the precursor solution is in the semidilute regime even at $p=0$.
In the gel state ($p_\mathrm{gel}<p\leq1$), $\Pi_\mathrm{mix}$ is constant, even when $p$ increases.
In general, the osmotic pressure is dependent and independent of the average molar mass in the dilute and semidilute regimes, respectively \cite{de1979scaling}.
Thus, the constant $\Pi_\mathrm{mix}$ in the gel state ($p_\mathrm{gel}<p\leq1$) indicates that polymer gels are always in the semidilute regime, because of the infinite molar mass of the polymer networks.\\

We can interpret $\Pi$ during the gelation process in the sol state ($0\leq p<p_\mathrm{gel}$) according to the universal EOS~(\ref{eq:EOS}).
By using $M$ and $c^{*}$ evaluated in Fig.~\ref{fig:measurement}(a) at each $p$, we changed the state variables (from $c$ and $\Pi$ to $\hat{c}$ and $\hat{\Pi}$), yielding the orange-filled circles in Fig.~\ref{fig:EOS}.
Remarkably, the osmotic pressure of the gelation process in the sol state ($p=0, 0.1, \dots, 0.5$) is described by the universal EOS~(\ref{eq:EOS}) of polymer solutions, although these systems continue to form multibranched polymer clusters.
Considering this finding in tandem with the semidilute scaling law observed in the gel state ($\Pi_{\mathrm{mix}}\propto c^{3\nu/(3\nu -1)}$), it is expected that $\hat{\Pi}_\mathrm{mix}$ in the gel state ($p_\mathrm{gel}<p\leq1$) conforms to the semidilute scaling law given by Eq.~(\ref{eq:scaling}) of \textit{linear}-polymer solutions [red line in Fig.~\ref{fig:result}(a)] with $K\simeq 1.1$.\\

Based on this expectation, we propose a universal EOS of osmotic pressure $\Pi_\mathrm{mix}$ for polymer gels as
\begin{equation}
K=\frac{\hat{\Pi}_\mathrm{mix}}{\hat{c}^{1/(3\nu-1)}}
\equiv \frac{M{c^{*}}^{1/(3\nu-1)} \Pi_\mathrm{mix}}{RTc^{\,3\nu/(3\nu-1)}},
\label{eq:scaling-gel}
\end{equation}
where $K\simeq 1.1$.
We note that $\hat{\Pi}_\mathrm{mix}\,\hat{c}^{1/(1-3\nu)}$ is finite, although both $\hat{c}\equiv c/c^{*}$ and $\hat{\Pi}_\mathrm{mix}\equiv \Pi_\mathrm{mix} M/(cRT)$ diverge to infinity, because gels correspond to infinite molar mass $M\to\infty$ and $c^{*}\to 0$.
In Fig.~\ref{fig:result}(b), we demonstrate that $\hat{\Pi}_\mathrm{mix}\,\hat{c}^{1/(1-3\nu)}$ converge to the universal value $K\simeq1.1$, which is independent of $p$ and $c$, after gelation ($p_\mathrm{gel}\leq p\leq 1$).
Therefore, in the inset of Fig.~\ref{fig:EOS}, the gel state is positioned at 
$(1/\hat{c},\hat{\Pi}_\mathrm{mix}\,\hat{c}^{1/(1-3\nu)}) \simeq (0, 1.1)$
(red star).
We obtained Fig.~\ref{fig:result}(b) by setting a constant value for $M{c^{*}}^{1/(3\nu-1)}$ and substituting $\Pi$ and $\Pi_\mathrm{mix}$ [shown in the inset of Fig.~\ref{fig:result}(a)] into Eqs.~(\ref{eq:scaling}) and (\ref{eq:scaling-gel}), respectively. 
(Further details are given in Sec.~S6 of Supplemental Material.)
This procedure demonstrates that we can determine $\Pi_\mathrm{mix}$ for any polymer gel by measuring the nonuniversal parameter $M{c^{*}}^{1/(3\nu-1)}$.\\

\textit{Concluding remarks}.---We experimentally measured the osmotic pressure of polymer gels throughout the gelation process.
We find that the universal EOS~(\ref{eq:EOS}) of osmotic pressure for \textit{linear}-polymer solutions describes the osmotic pressure throughout the gelation process involving both the sol and gel states (Fig.~\ref{fig:EOS}).
In the sol state, both $\hat{\Pi}$ and $\hat{c}$ continuously increase according to the universal EOS~(\ref{eq:EOS}) with an increase in the average molar mass (orange-filled circles in Fig.~\ref{fig:EOS}).
In the gel state, the osmotic pressure of the polymer gels is universally governed by the semidilute scaling law~(\ref{eq:scaling-gel}) [red star in Fig.~\ref{fig:EOS} and Fig.~4(b)].
Here, both $\hat{\Pi}$ and $\hat{c}$ diverge to infinity, because the gel state corresponds to the average molar mass $M\to\infty$ and the overlap concentration $c^{*}\to 0$.
In addition, we have demonstrated that Eq.~(\ref{eq:scaling-gel}) enables the determination of $\Pi_\mathrm{mix}$ for any polymer gel by measuring a nonuniversal parameter $M{c^{*}}^{1/(3\nu-1)}$.

Our results provide a new system for demonstrating universality in statistical physics because the universal EOS~(\ref{eq:EOS}) relates to the $O$($n$)-symmetric universal classes \cite{pelissetto2002critical}.
In addition, a universal EOS and universal thermodynamics are of great interest to those studying strongly correlated fermions \cite{fermion}.
Our findings can stimulate research in these fields.

\begin{acknowledgments}
We would like to thank Masao~Doi, Yuichi~Masubuchi, Takashi~Uneyama, and Xiang~Li for their useful comments.
This work was supported by the Japan Society for the Promotion of Science (JSPS) through the Grants-in-Aid for 
Early Career Scientists Grant No. 19K14672 to N.S.,
Scientific Research (B) Grant No. 18H02027 to T.S.,
and Scientific Research (S) Grant No. 16H06312 to U.C. 
This work was also supported by the Japan Science and Technology Agency (JST) CREST Grant No JPMJCR1992 to T.S. and COI Grant No. JPMJCE1304 to U.C.
\end{acknowledgments}

\bibliographystyle{apsrev}

\end{document}


\title{Supplemental Material for:
``Universal Equation of State Describes Osmotic Pressure throughout Gelation Process''}
\author{Takashi~Yasuda}
\thanks{These authors contributed equally: T.~Yasuda, N.~Sakumichi}
\affiliation{Department of Bioengineering, The University of Tokyo, 7-3-1 Hongo, Bunkyo-ku, Tokyo, Japan.}
\author{Naoyuki~Sakumichi}
\email[Correspondence should be addressed to N.~Sakumichi or T.~Sakai: ]{sakumichi@tetrapod.t.u-tokyo.ac.jp}
\affiliation{Department of Bioengineering, The University of Tokyo, 7-3-1 Hongo, Bunkyo-ku, Tokyo, Japan.}
\author{Ung-il~Chung}
\affiliation{Department of Bioengineering, The University of Tokyo, 7-3-1 Hongo, Bunkyo-ku, Tokyo, Japan.}
\author{Takamasa~Sakai}
\email{sakai@tetrapod.t.u-tokyo.ac.jp}
\affiliation{Department of Bioengineering, The University of Tokyo, 7-3-1 Hongo, Bunkyo-ku, Tokyo, Japan.}
\date{\today}

\maketitle
\noindent

\section{Definition of Polymer Concentration}
\label{sec:Concentration}

In the main text, we adopted a different definition of polymer concentration ($c_{0}$ and $c$) than that used conventionally ($\tilde{c}_{0}$ and $\tilde{c}$) in order to extend the universality of the equation of state (EOS) of osmotic pressure to higher concentrations.
In this section, we demonstrate the validity of this claim by comparing the two EOSs for linear-polymer solutions determined using each definition.

The conventional polymer concentration is defined as
\begin{equation}
\tilde{c}=\frac{W_\mathrm{polymer}}{V_\mathrm{solution}},
\label{eq:solution}
\end{equation}
where $W_\mathrm{polymer}$ and $V_\mathrm{solution}$ are the polymer weight and solution volume, respectively.
We note that $V_\mathrm{solution}$ includes not only the solvent volume $V_\mathrm{solvent}$, but also the polymer volume $V_\mathrm{polymer}$ as $V_\mathrm{solution}\simeq V_\mathrm{solvent}+V_\mathrm{polymer}$.
For the EOS to be universal, the polymer chains must be excessively thin, and their volumes must be negligible \cite{de1979scaling};
the universality is affected by $V_\mathrm{polymer}$ as the polymer concentration increases.
Thus, we adopt the following definition:
\begin{equation}
c=\frac{W_\mathrm{polymer}}{V_\mathrm{solvent}}.
\label{eq:solvent}
\end{equation}
This definition in Eq.~(\ref{eq:solvent}) is similar to the molecule volume consideration in the van der Waals EOS for a real gas.
In other words, the extent of polymer chain motion is considered to be $V_\mathrm{solvent}$, which is the total solution volume $V_\mathrm{solution}$ minus its own volume $V_\mathrm{polymer}$.

\begin{figure}[t!]
\centering
\includegraphics[width=\linewidth]{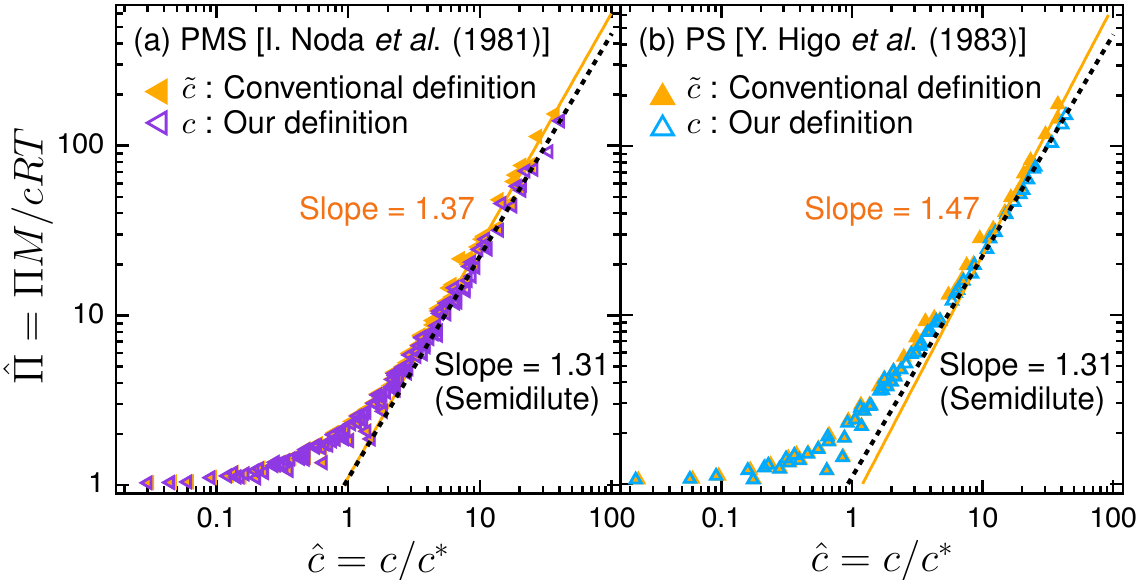}
\caption{
Comparison of EOSs using the different definitions of polymer concentration given in Eqs.~(\ref{eq:solution}) and (\ref{eq:solvent}) for (a) poly($\alpha$-methylstyrene) \cite{noda1981thermodynamic} and (b) poly(styrene) \cite{higo1983osmotic} in toluene solutions.
We use $\tilde{c}$ and $c$ to plot the filled and unfilled triangles, respectively.
In the semidilute regime ($c^{*}<c$), the universal EOS is asymptotic to the scaling law \cite{de1979scaling} as
$\hat{\Pi} \propto \hat{c}^{1/(3\nu -1)}$ (Eq.~(3) in the main text),
where $1/(3\nu -1)\simeq 1.31$ for a good solvent ($\nu\simeq 0.588$).
}
\label{fig:concentration}
\end{figure}

Figure~\ref{fig:concentration} compares the two EOSs determined using the definitions in Eqs.~(\ref{eq:solution}) and (\ref{eq:solvent}) for the two linear polymer solutions of (a) poly($\alpha$-methylstyrene) (PMS) \cite{noda1981thermodynamic} and (b) poly(styrene) (PS) \cite{higo1983osmotic}.
Adopting the definition given by Eq.~(\ref{eq:solution}), the scaling slopes in the semidilute regime are $1.37$ and $1.47$ in the PMS and PS solutions, respectively.
These slopes are greater than those in the theoretical prediction $1/(3\nu -1)\simeq 1.31$ for a good solvent ($\nu\simeq 0.588$) \cite{de1979scaling} because of the influence of the polymer volume at high concentrations.
However, adopting the definition in Eq.~(\ref{eq:solvent}),
the scaling slopes in the semidilute regime for both the PMS and PS solutions asymptotically approach the theoretical prediction ($1.31$) as $c$ increases.
This emphasizes the universality of the polymer solutions in higher concentrations when adopting the concentration definition given in Eq.~(\ref{eq:solvent}).
In the main text, therefore, we adopted Eq.~(\ref{eq:solvent})
as the definition of the polymer concentration to analyze the universal scaling function.

\section{Limitations of the measurement of osmotic pressure}

In the experiment, the measurement of osmotic pressure via osmotic deswelling faces several limitations.
Precursor-solution samples ($p=0$) of higher concentrations ($c_0>120$ g$/$L) were not measured because of our limited knowledge of the external polymer (poly(vinylpyrrolidone)(PVP)) solutions $\Pi_{\mathrm{ext}}$.
Sol samples ($0\leq p<p_\mathrm{gel}$) near their gelation points $p_\mathrm{gel}$ could not be measured because of their viscosity divergences.
Gel samples with low degrees of connectivity ($p_{\mathrm{gel}} < p < 0.7$) were difficult to characterize for the following two reasons:
(i) the gels lost their shapes and weights because of their low elasticity, and (ii) we cannot uniquely determine the equilibrium state in the present osmotic pressure measurement because of the outflow of small polymer clusters.
Gel samples ($p=1$) at low concentrations ($c_{0}<40$ g$/$L) were excluded from the analysis because these samples became heterogeneous as a result of phase separation.

\begin{figure}[t!]
\centering
\includegraphics[width=\linewidth]{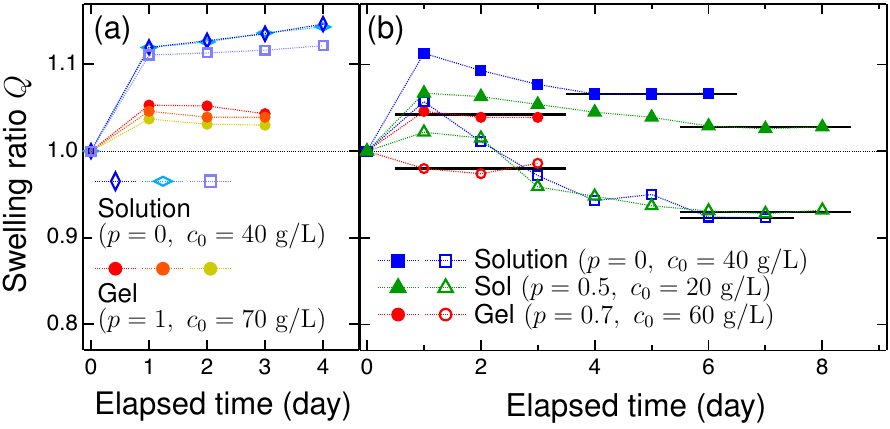}
\caption{
Time courses of swelling ratio $Q$ in the swelling experiments (see Fig.~3 in the main text).
We used solution ($p=0$), sol ($0<p<p_{\mathrm{gel}}$), and gel ($p \geq p_{\mathrm{gel}}$) samples whose precursors were $M=10$ kg$/$mol.
(a) Reproducibility of $Q$ for the solution and gel samples in the external polymer (PVP) solutions on $c_\mathrm{ext}=70$ g$/$L.
For each sample, the three representative experiments agreed with each other, indicating high reproducibility.
(b) Relaxation to swelling equilibrium.
For the solution samples, we set $c_\mathrm{ext}=75$ g$/$L (blue filled squares) and $c_\mathrm{ext}=85$ g$/$L (blue open squares).
For the sol samples, we set $c_\mathrm{ext}=35$ g$/$L (green filled triangles) and $c_\mathrm{ext}=40$ g$/$L (green open triangles).
For the gel samples, we set $c_\mathrm{ext}=70$ g$/$L (red filled circles) and $c_\mathrm{ext}=75$ g$/$L (red open circles).
}
\label{fig:swelling}
\end{figure}

\section{Osmotic pressure of PVP solutions}

In the main text, we measured the osmotic pressure of the samples using
controlled aqueous PVP solutions.
According to Vink \cite{vink1971precision}, the osmotic pressure $\Pi_{\mathrm{ext}}$ of an aqueous PVP solution is given as
\begin{equation}
\Pi_{\mathrm{ext}} = 21.27 c_{\mathrm {ext}} + 1.63 {c_{\mathrm {ext}}}^2 +0.0166 {c_{\mathrm {ext}}}^3,
\label{eq:PVP}
\end{equation}
for the PVP concentration $c_\mathrm{ext} \leq 200$ g$/$L.
Equation~(\ref{eq:PVP}) was precisely determined using an osmometer with an organic liquid manometer, mercury manometer, and manostat.
We note that $c_{\mathrm{ext}}$ in Eq.~(\ref{eq:PVP}) is defined as the polymer weight divided by the solution volume, i.e., Eq.~(\ref{eq:solution}), which differs from the definition of the polymer concentration ($c$ and $c_0$) used in the main text, i.e., Eq.~(\ref{eq:solvent}) (see Supplementary Material, Sec.~\ref{sec:Concentration}).

\section{Achievement of swelling equilibrium}

To confirm the reproducibility of the swelling experiment described in the main text, we show the swelling ratio $Q$ as a function of the elapsed time for the solution ($p=0$) and gel ($p=1$) samples in Fig.~\ref{fig:swelling}(a).
We immersed each sample (with or without the micro-dialyzer) in the external polymer (PVP) solution at a certain concentration $c_\mathrm{ext}$ (see Fig.~3 in the main text).
Each sample changed in weight ($w_0 \to w$) and concentration ($c_0 \to c$) from the initial to equilibrium states.
We determined the equilibrium swelling ratio as $Q=w/w_0$.

To ensure that each sample achieved the equilibrium state in the swelling experiment, we show the relaxation processes for the solutions ($p=0$), sol ($0<p<p_{\mathrm{gel}}$), and gel ($p_{\mathrm{gel}}<p\leq 1$) samples 
in Fig.~\ref{fig:swelling}(b).
For each sample, we achieved both the swelling (filled symbols) and deswelling (unfilled symbols) conditions, by adjusting the concentration of the external solutions $c_\mathrm{ext}$.
To achieve the swelling equilibrium (horizontal black lines), one week was required for the solution and sol samples, and a few days were required for the gel samples.
Here, we determined the swelling equilibrium as the point at which the swelling ratio $Q$ remained constant for three days.

\begin{figure}[t!]
\centering
\includegraphics[width=8.2cm]{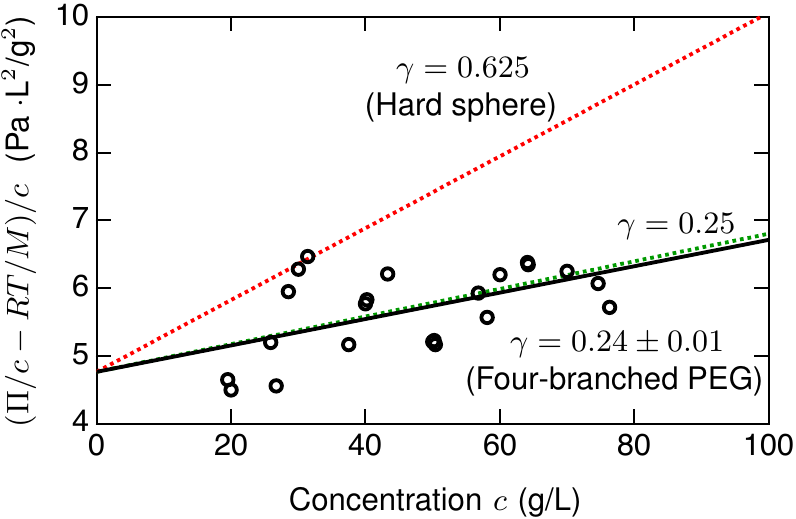}
\caption{
Concentration dependence of $(\Pi/c-RT/M)/c$ for four-branched polymer (PEG) solutions (black circles) with $M = 8.8(1)$ kg$/$mol and $c^* =58.5(4)$ g$/$L.
Here, the values in parentheses represent the standard error estimated from the blue circles and line ($p=0$) in Fig.~3(a) in the main text.
The black line represents the least-squares fit.
According to Eq.~(\ref{eq:virial}), the black line gives $\gamma$ of the four-branched polymer (PEG) solutions as $\gamma=0.24(1)$.
For comparison, we show the slopes corresponding to $\gamma = 0.25$ for linear-polymer solutions \cite{flory1953principles} and $\gamma = 0.625$ for hard spheres \cite{carnahan1969equation}.
}
\label{fig:thirdvirial}
\end{figure}

\section{Third virial coefficient of four-branched polymer solution}

We evaluated the third virial coefficient based on the $c$-dependence of $\Pi$ in the four-branched polymer (PEG) solution (i.e., the precursor solutions ($p=0$) in Fig.~4(a) in the main text).
In terms of the universal EOS, the third virial coefficient corresponds to the dimensionless virial ratio $\gamma$ defined in Eq.~(2) in the main text.
From the virial expansion up to the third terms (Eq.~(2) in the main text), we have 
\begin{equation}
\left( \frac{\Pi}{c} - \frac{RT}{M} \right) \frac{1}{c}= 
\frac{RT}{Mc^{*}}
\left(1 + \frac{c\gamma}{c^{*}}
\right) .
\label{eq:virial}
\end{equation}
In Fig.~3(a) in the main text, we obtain $M=8.9(1)$ kg$/$mol and $c^*=58.5(4)$ g$/$L at $T=298$ K, where the values in parentheses represent the standard errors.
Thus, the $c$-dependence of $\left(\Pi/c - RT/M \right)/c$ gives $\gamma$ of the four-branched polymer (PEG) solutions, as shown in Fig.~\ref{fig:thirdvirial}.
This analysis yields $\gamma=0.24(1)$, which is consistent with $\gamma \simeq 0.25$ for \textit{linear}-polymer solutions \cite{flory1953principles} within the error bounds.
The consistency of $\gamma$ between the linear and four-branched polymer solutions indicates that the effect of a small number of branches on osmotic pressure is negligibly small in terms of the universal EOS.

\begin{figure}[t!]
\centering
\includegraphics[width=8.2cm]{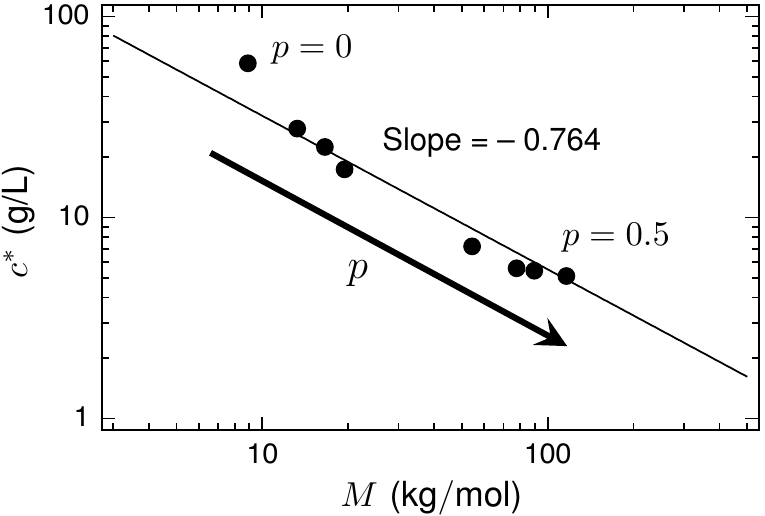}
\caption{
Relationship between the overlap concentration $c^{*}$ and the number average molar mass $M$ during the gelation process in the sol state ($0\leq p<p_\mathrm{gel}$).
The precursors are $M=10$ kg$/$mol on $c_0=20$ g$/$L for $p=0, 0.1, 0.2, 0.25, 0.3, 0.35, 0.4, 0.5$.
This result agrees with the scaling prediction in Eq.~(\ref{sup:eq:cStar-M}) with $1-3\nu\simeq -0.764$ for a good solvent ($\nu\simeq 0.588$).
}
\label{fig:cstar}
\end{figure}

\section{$M$ and $c^{*}$ during the gelation process}

We experimentally determined the average molar mass $M$ and the overlap concentration $c^{*}$ during the gelation process to obtain the universal scaling parameters $\hat{c}$ and $\hat{\Pi}$ in Figs.~1 and 4(b) in the main text.
When gelation proceeds at a constant concentration $c_{0}$, $M$ increases and the corresponding $c^{*}$ decreases.
Assuming the scaling relation of the gyration radius $R_{\mathrm g}$ to be $\langle {R_{\mathrm g}}^{2}\rangle \propto M^{2\nu}$ \cite{de1979scaling}, we have
\begin{equation}
c^{*}\equiv\frac{M}{4\pi^{\frac{3}{2}}\langle {R_{\mathrm g}}^{2}\rangle^{\frac{3}{2}}N_{\mathrm A}\Psi^{*}} \propto M^{1-3\nu},
\label{sup:eq:cStar-M}
\end{equation}
during the gelation process ($0\leq p \leq1$) if the interpenetration factor $\Psi^{*}$ is constant as $p$ increases.

\begin{figure}[t!]
\centering
\includegraphics[width=\linewidth]{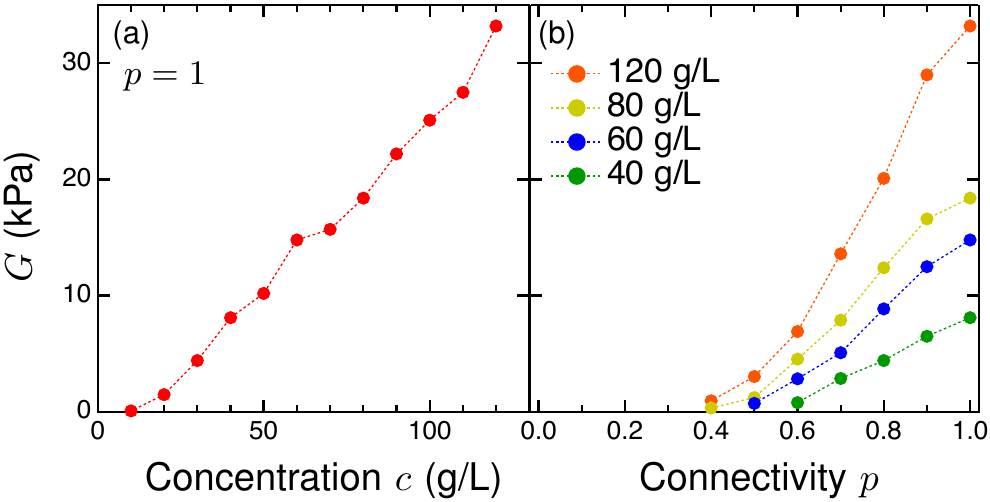}
\caption{
Shear moduli $G$ of gel samples with the molar mass of precursors $M=10$ kg$/$mol.
(a) Dependence of $G$ on the polymer concentration $c$ at a constant connectivity $p=1$.
We used this data to determine the $c$-dependence of $\Pi_\mathrm{mix}$ shown in the main panel of Fig.~4(a).
(b) Dependence of $G$ on connectivity ($p = 2s = 0.4, 0.5, 0.6, 0.7, 0.8, 0.9, 1$) for various polymer concentrations ($c = 40, 60, 80, 120$ g$/$L).
We used this data to determine the $p$-dependence of $\Pi_\mathrm{mix}$ shown in the inset of Fig.~4(a).
}
\label{fig:shearmodulus}
\end{figure}

\begin{figure}[t!]
\centering
\includegraphics[width=\linewidth]{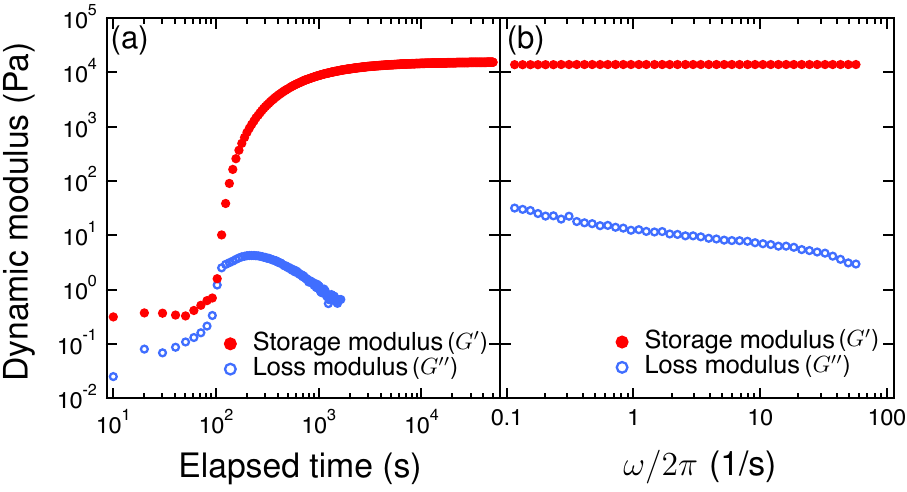}
\caption{
Storage modulus $G'$ and loss modulus $G''$ measured by a dynamic shear rheometer at $298$~K.
The sample was prepared with the molar mass of precursors $M=10$ kg$/$mol, the concentration $c=70$ g$/$L, and the connectivity $p=1$.
(a) Time courses of $G'$ and $G''$ during the chemical reaction of the cross-end coupling.
The applied strain and frequency are $1 \%$ and $1$~Hz, respectively.
(b) Frequency dependences of $G'$ and $G''$ after completion of the chemical reaction ($t\gg t_{\mathrm{gel}}$).
}
\label{fig:rheometer}
\end{figure}

To confirm $M$ and $c^{*}$ in the sol state ($0\leq p <p_\mathrm{gel}$) for $c_{0}=20$ g$/$L, which were experimentally obtained in Fig.~3(a) in the main text, we compare the results with the corresponding scaling law of Eq.~(\ref{sup:eq:cStar-M}).
Each $M$ and $c^{*}$ was used to yield the orange filled circles shown in Fig.~(1) in the main text.
Figure~\ref{fig:cstar} shows that the obtained $M$ and $c^{*}$ in the sol state (black circles) were consistent with the scaling prediction (black line) of Eq.~(\ref{sup:eq:cStar-M}).
Here, the deviation at $p=0$ occurred because gelation proceeds in the dilute regime ($c_{0}<c^{*}$) with an increase in $\Psi^{*}$.
After the system reaches the semidilute regime ($c>c^{}{*}$), the scaling law of Eq.~(\ref{sup:eq:cStar-M}) holds with a constant $\Psi^{*}$ as $p$ increases.

To evaluate $\hat{\Pi}/\hat{c}^{1.31}$ in Fig.~4(b) in the main text, we substituted $M{c^{*}}^{1/(3\nu-1)}= 1600$ $(\mathrm{kg}/\mathrm{mol})(\mathrm{g}/\mathrm{L})^{1/(3\nu-1)}$ into Eqs.~(3) and (4) throughout the gelation process for $c_{0}= 40, 60, 80, 120$ g$/$L.
Here, we assumed that $M{c^{*}}^{1/(3\nu-1)}$ is constant during the gelation process in the semidilute regime.
This is because $\Pi$ and $\Pi_\mathrm{mix}$ are constant in the semidilute regime (see orange squares for $c_{0}=120$ g$/$L in the inset of Fig.~4(a) in the main text) and the other parameters ($R$, $T$, $c$, and $K \simeq 1.1$) are also constant in Eqs.~(3) and (4) in the main text.

\section{Measurement of shear modulus}

To determine $\Pi_\mathrm{mix}$ shown in Fig.~4(a) in the main text,
we measured the dependences of the shear moduli $G$ of the gel samples on the concentration $c$ (Fig.~\ref{fig:shearmodulus}(a)) and connectivity $p$ (Fig.~\ref{fig:shearmodulus}(b)) by using dynamic viscoelasticity measurements.
The tetra-PEG MA and tetra-PEG SH aqueous solutions were mixed and poured into the gap of the double cylinder of the dynamic shear rheometer (MCR 301 and 302, Anton Paar, Graz, Austria).
Subsequently, we cyclically sheared the samples in the gap between the inner cylinder and outer cup.
The stress--strain response yielded the storage moduli $G'$ and loss moduli $G''$ of the samples.

Figure~\ref{fig:rheometer}(a) shows the typical time course of $G'$ and $G''$.
After $G'$ reached equilibrium, which corresponds to the completion of the chemical reaction of the MA and SH functional groups, we measured the frequency dependence of $G'$ and $G''$ (Fig.~\ref{fig:rheometer}(b)).
The equilibrium shear modulus is given by
\begin{equation}
G=\lim_{\omega\to 0} G'(\omega).
\end{equation}
Figure~\ref{fig:rheometer}(b) shows that $G'(\omega)$ is independent of the frequency ($\omega/2\pi$) below $100$~Hz.
Thus, we regard $G'(\omega)$ at $1$ Hz as the shear modulus $G$.